\documentclass[floatfix, ApJL,twocolumn]{aastex631}

\usepackage{amssymb}
\usepackage{amsmath}
\usepackage{aas_macros}
\usepackage{color,units}
\usepackage{xspace}
\usepackage{subfigure}
\usepackage{comment}
\usepackage{xcolor}
\usepackage{bm}
\usepackage{dcolumn}
\usepackage{hyperref}
\usepackage{enumitem} 
\usepackage{array}
\usepackage{nicematrix}
\usepackage{hhline}
\usepackage{graphicx}
\usepackage{multirow}
\usepackage[normalem]{ulem}
\usepackage{soul}
\usepackage[math]{cellspace}
\usepackage{nicematrix} 

\BeforeBegin{NiceTabular}{\let\begin\BeginEnvironment\let\end\EndEnvironment}
\BeforeBegin{NiceArray}{\let\begin\BeginEnvironment}
\BeforeBegin{NiceMatrix}{\let\begin\BeginEnvironment}

\newcommand{\SPA}{School of Physics and Astronomy, Monash University, Clayton VIC 3800, Australia}
\newcommand{\OzGravMonash}{OzGrav: The ARC Centre of Excellence for Gravitational Wave Discovery, Clayton VIC 3800, Australia}

\newcommand{\AMNH}{Department of Astrophysics, American Museum of Natural History, New York, NY 10024, USA}
\newcommand{\CCA}{Center for Computational Astrophysics, Flatiron Institute, New York, NY 10010, USA}
\newcommand{\CUNY}{Graduate Center, City University of New York, 365 5th Avenue, New York, NY 10016, USA}
\newcommand{\BMCC}{Department of Science, BMCC, City University of New York, New York, NY 10007, USA}

\newcommand{\imrxhm}{{\sc \texttt{IMRPhenomXPHM}}\xspace}
\newcommand{\fancytext}[1]{{\relax\ifmmode#1\else $#1$\fi}\xspace}
\newcommand{\mathcmd}[1]{{\sc \relax\ifmmode#1\else $#1$\fi}\xspace}

\newcommand{\Nt}{\mathcmd{\cal{N}_{\rm t} }}

\newcommand{\lowRho}{$\overline{\rho} < \unit[10^{-11}]{gcm^{-3}}$}
\newcommand{\highRho}{$\overline{\rho} > \unit[10^{-11}]{gcm^{-3}}$}

\newcommand{\ancientTau}{$\overline{\tau}_{\rm AGN} \gg \unit[5]{Myr}$}
\newcommand{\oldTau}{$\overline{\tau}_{\rm AGN} \ga \unit[5]{Myr}$}
\newcommand{\midTau}{$\overline{\tau}_{\rm AGN} \sim \unit[0.5-5]{Myr}$}
\newcommand{\youngTau}{$\overline{\tau}_{\rm AGN} \la \unit[1]{Myr}$}

\newcommand{\Lalign}{$\vec{L} \parallel \vec{J}_{\rm{AGN}}$}
\newcommand{\Liso}{$\vec{L}$ Isotropic}

\newcommand{\xJalign}{$\vec{\chi}_1 \parallel \vec{J}_{\rm{AGN}}$}
\newcommand{\xLalign}{$\vec{\chi}_1 \parallel \vec{L}$}
\newcommand{\xLiso}{$\vec{\chi}_1$ Isotropic}

\newcommand{\xxalign}{${\vec{\chi}_2 \parallel \vec{\chi}_1}$}
\newcommand{\xxiso}{$\vec{\chi}_2$ Isotropic}

\newcommand{\tdampSmall}{$t_{\rm damp} < t_{\rm enc}$}
\newcommand{\tdampBig}{$t_{\rm damp} > t_{\rm enc}$}

\newcommand{\bigSig}{\large$\bm{\sigma_1 > 1}$}
\newcommand{\lilSig}{\large$\bm{\sigma_1 \leq 1}$}
\newcommand{\bigRSig}{\large$\bm{\sigma_{12} > 1}$}
\newcommand{\lilRSig}{\large$\bm{\sigma_{12} \leq 1}$}

\newcommand{\lilDm}{$\Delta m \ll 0.01\, m_i$}
\newcommand{\bigDm}{$\Delta m \ga 0.01\, m_i$}
\newcommand{\edDm}{$\Delta m \sim 0.01\, m_i$}

\newcommand{\bigE}{$\epsilon_{\rm gas}\rightarrow1$}
\newcommand{\midE}{$0<\epsilon_{\rm gas}<1$}
\newcommand{\lowE}{$\epsilon_{\rm gas}\ll 1$}

\newcommand{\topL}{(a) \textit{Old, Dense AGN}}
\newcommand{\topR}{(b) \textit{Mid-Aged, Dense AGN}}
\newcommand{\botL}{(c) \textit{Old, Dilute AGN}}
\newcommand{\botR}{(d) \textit{Young, Dilute AGN}}

\newcommand{\code}[1]{{\texttt{#1}}\xspace}
\newcommand{\bilby}{\code{bilby}}
\newcommand{\parallelbilby}{\code{parallel-bilby}}
\newcommand{\gwpop}{\code{GWPopulation}}
\newcommand{\bilbypipe}{\code{bilby-pipe}}
\newcommand{\dynesty}{\code{dynesty}}
\newcommand{\gwpy}{\code{GWpy}}
\newcommand{\pvec}[1]{\vec{#1}\mkern2mu\vphantom{#1}}

\begin{document}

\title{Measuring the properties of active galactic nuclei disks with gravitational waves}


\shortauthors{Vajpeyi et al.}
\author{Avi Vajpeyi}
\email{avi.vajpeyi@monash.edu}
\affiliation{\SPA}
\affiliation{\OzGravMonash}

\author{Eric Thrane}
\affiliation{\SPA}
\affiliation{\OzGravMonash}

\author{Rory Smith}
\affiliation{\SPA}
\affiliation{\OzGravMonash}

\author{Barry McKernan} 
\affiliation{\AMNH}
\affiliation{\CCA}
\affiliation{\CUNY}
\affiliation{\BMCC}

\author{K.E. Saavik Ford}
\affiliation{\AMNH}
\affiliation{\CCA}
\affiliation{\CUNY}
\affiliation{\BMCC}

\begin{abstract}
Active galactic nuclei (AGN) are promising environments for the assembly of merging binary black hole (BBH) systems.
Interest in AGNs as nurseries for merging BBH is rising following the detection of gravitational waves from a BBH system from the purported pair-instability mass gap, most notably, GW190521.
Active galactic nuclei have also been invoked to explain the formation of the high-mass-ratio system, GW190814.
We draw on simulations of BBH systems in AGN to propose a phenomenological model for the distribution of black hole spins of merging binaries in AGN disks.
The model incorporates distinct features that make the AGN channel potentially distinguishable from other channels, such as assembly in the field and in globular clusters.
The model parameters can be mapped heuristically to the age and density of AGN disks.
We estimate the extent to which different populations of mergers in AGNs can be distinguished.
If most merging black holes are assembled in AGNs, future gravitational-wave observations may provide insights into the dynamics of AGN disks.
\end{abstract}

\section{Introduction}
Gravitational-waves from the mergers of binary black hole (BBH) systems have recently transformed astronomy.
However, the astrophysical origins of these events are still uncertain.
There are two main proposed astrophysical pathways to the mergers: (i) isolated binary evolution via mass transfer, including a common envelope phase in galactic fields; (ii) dynamical formation in dense environments. 
Each pathway is associated with different distributions of black-hole spin \citep{2010CQGra..27k4007M,Stevenson,Fishbach:2017dwv,spin_population_models,Wysocki2019} and binary eccentricity \citep{Rodriguez2018,Samsing2018,eccentricity,gwtc_eccentricity,gwtc2_eccentricity,eccentric_implications,Gayathri2020}.
Measuring BBH spins and eccentricity with gravitational waves can therefore be used to determine how and where BBH are assembled \citep{o2_pop,Abbott:2021:ApJL}. 

Active Galactic nuclei (AGNs) are expected to contain a dense population of stars and stellar remnants, such as stellar-origin black holes (BHs) \citep{1993ApJ...408..496M,2000ApJ...545..847M,2018Natur.556...70H,2018MNRAS.478.4030G}.
Binary black hole systems can form via close encounters in this dynamically ``hot'' environment but are often rapidly ``ionized'' via tertiary encounters \citep{2016ApJ...831..187A,2019MNRAS.488...47F}.
The dense nuclear population and AGN gas disks (when present) can interact, resulting in an embedded population of stars and BH within the disk.
These embedded objects can weakly perturb the surface-density profile of the gas disk, resulting in gas torques within AGN disk that allow for Type I (non gap-opening) migration of the embedded objects \citep{2012MNRAS.425..460M}. 
Differential migration rates of the objects encourage binary formation, leading to compact binary mergers detectable with LIGO-Virgo \citep{2014MNRAS.441..900M,2017ApJ...835..165B,2017MNRAS.464..946S,2021arXiv211010838W, ligo, virgo}.
Kicked merger products are generally retained by the deep potential well, allowing for hierarchical BBH mergers.
If most mergers observed by LIGO--Virgo were assembled in an AGN disk, it may be possible to reverse-engineer conditions beneath the AGN photosphere \citep{2018ApJ...866...66M,2021arXiv211010838W}. 

The gaseous disk in AGNs likely serve to align (to varying degrees) both black-hole spin vectors\footnote{The subscript $1$ refers to the primary, more massive black hole while the $2$ refers to the less massive secondary.} $\vec\chi_{1,2}$ and the binaries' orbital angular momentum vectors $\vec{L}$---depending on the density and age of the disk \citep{2007ApJ...661L.147B}.
On the other hand, tertiary encounters with binaries in the disk tend to misalign $\vec\chi_{1,2}$ relative to $\vec{L}$---depending on the timescales of encounters~\citep{Liu:2017:ApJL, Tagawa:2020:ApJ}. 
The competing effects of the gaseous disk and dynamical encounters on BBHs in AGNs determine the distribution of BBH spin orientations. On the other hand, binaries born in the field have $\vec\chi_{1,2}$ nearly aligned to $\vec{L}$ with a small spread due to supernova kicks \citep{2000ApJ...541..319K, 2010CQGra..27k4007M, 2013ApJ...779...72D, 2018MNRAS.474.2959G, 2017PASA...34...58E, 2020arXiv200411866O}. 
Finally, dynamically assembled binaries (e.g., in globular clusters) exhibit no correlation between the $\vec{L}$ and $\vec\chi_{1,2}$. 
In this paper, we propose a phenomenological model for the distribution of black-hole spins in AGNs to capture the salient features predicted from theoretical modeling. 

The remainder of this paper is organised as follows.
In Section~\ref{sec:agn}, we review the spin-orientations of BBH in AGNs.
Section~\ref{sec:model} presents a phenomenological model describing AGN BBH spins-orientations.
In Section~\ref{sec:results} we present the results of a simulated study, demonstrating the model's effectiveness. 

\section{Spin properties of binary black holes at formation}\label{sec:agn}

\begin{table*}
\centering
\caption{
\textbf{The phenomenology of black holes spin for binaries merging in AGNs.}
The two rows correspond to different values of the population parameter $\sigma_1$ defined in Section~\ref{sec:model}, which controls the effective density of the AGN.
The two columns correspond to different values of the population parameter $\sigma_{12}$ defined in Section~\ref{sec:model}, which controls the effective age of the AGN.
Each cell is divided into two.
The cell's left-hand side describes the AGN's properties while the right-hand side describes the distribution of black hole spins \textit{at the time of formation}.
}
\label{tab:astrophysical-imp}
\def\arraystretch{1.3}
\setlength\cellspacetoplimit{5pt}
\setlength\cellspacebottomlimit{5pt}
\setlength{\tabcolsep}{1.2em}
\begin{NiceTabular}{l|l|l|l|l}
                     & \Block{1-2}{\lilRSig} & & \Block{1-2}{\bigRSig} &  \\ \hline
\Block{7-1}{\lilSig} & \Block{1-2}{\topL}    & & \Block{1-2}{\topR} &     \\
                     & \oldTau      & \Lalign  & \midTau      & \Lalign   \\
                     & \highRho     & \xLalign & \highRho     & \xLalign  \\
                     & \tdampSmall  & \xxalign & \tdampSmall  & \xxiso    \\
                     & \bigDm       &          & \lilDm       &           \\
                     & \bigE        &          & \midE        &           \\
                     & \Block{1-2}{}&          & \Block{1-2}{}&           \\ \hline
\Block{7-1}{\bigSig} & \Block{1-2}{\botL}    & & \Block{1-2}{\botR} &    \\
                     & \oldTau      & \Liso    & \youngTau    & \Liso     \\
                     & \lowRho      & \xJalign & \lowRho      & \xLiso    \\
                     & \tdampBig    & \xxalign & \tdampBig    & \xxiso    \\
                     & \edDm        &          & \lilDm       &           \\
                     & \midE        &          & \lowE        &           \\
                     & \Block{1-2}{}&          & \Block{1-2}{}&           
\end{NiceTabular}
\end{table*}


Modeling AGNs is challenging due to the interplay between gas dynamics, scattering binaries, and feedback from the central supermassive BH.
Commonly used disk models span wide ranges of disk density and geometry \citep[see, e.g.,][]{2003MNRAS.341..501S,2005ApJ...630..167T}, and provide broad estimates for merger rates in AGNs, \citep[e.g.,][]{2018ApJ...866...66M,2020A&A...638A.119G,Tagawa:2020:ApJ}.
However, we have qualitative predictions for the spin-orientation population properties of merging BBH systems using Monte-Carlo and $N$-body simulations to identify key features. 
Some of these predictions are tabulated in Table~\ref{tab:astrophysical-imp}. 

In this section, we discuss three predictions for the spin distribution of merging BBH at the time the binary is formed\footnote{We assume BBH merger timescale are of the order $\unit[0.1-1]{Myr}$ \citep{Baruteau:2011:ApJ, Tagawa:2020:ApJ, 2020ApJ...903..133S, 2020ApJ...901L..34Y, 2020MNRAS.498.4088M}.}.
The spin vectors subsequently evolve through general relativistic precession of the orbital plane.
Nonetheless, the orientation of the spin vectors at the merger contains information about the orientation at formation.
We describe the phenomenology of the spin vectors at the formation and then discuss later the resulting phenomenology at the time of the merger.

\begin{enumerate}[label=(\roman*)]
\item 
\textbf{Gas accretion torques BH spins to align with the disk.} 
The BH embedded in the AGN disk early in the disk lifetime should have isotropically distributed spins at formation~\citep{2016LNP...905..205M, Tagawa:2020:ApJ}. 
As the BH of mass $m$ migrate in the disk, they accrete $\Delta m$ disk-gas mass, resulting in a torque pointing into the plane of the AGN disk. 
The magnitude of the torque on BH spins depends on $\Delta m$ \citep{2007ApJ...661L.147B} and can be summarised as follows.

Dense gas disks can torque initially randomly oriented BH spins into alignment with the disk angular momentum in $< \unit[5]{Myr}$ \citep[e.g.][]{2020MNRAS.498.4088M}, assuming an Eddington-limited accretion rate. 
\citet{2020MNRAS.499.2608F} find that a critical density of\ \highRho\ is required to capture many orbiters into the disk over a modest AGN lifetime. 
So, here we assume that gas disks with densities\ \lowRho\ need to live $> \unit[5]{Myr}$ both to capture enough BH over their lifetimes to make a significant contribution to the BBH merger rate and to torque BH spins into alignment with the disk.

In sufficiently long-lived, dense disks (Table~\ref{tab:astrophysical-imp}a), fully embedded BH accrete more than $1-10\%$ of their initial mass \bigDm. 
The resultant torque from gas accretion onto the embedded BH reorients the BH spin vector $\vec{\chi}$ to align with the angular momentum vector for the AGN disk $\vec{J}_\text{AGN}$, \citep[e.g.,][]{2007ApJ...661L.147B}. 

Alternatively, for dilute disks (\lowRho, Table~\ref{tab:astrophysical-imp}cd), \lilDm\ implies that the BH are not torqued into alignment with the disk in \midTau~\citep{2007ApJ...661L.147B}. 
Similarly, if AGN disks are dense but typically short-lived (\youngTau), this effect is weaker (Table~\ref{tab:astrophysical-imp}ac).

Note that the details of accretion onto objects embedded in AGN disks is subject to much uncertainty.
Feedback, turbulence, and interactions can alter gas flow dynamics inside the BBH Hill sphere, possibly inhibiting a high accretion rate on component BH, limiting the average torque magnitude~\citep{Hankla:2020:ApJ}.
Lower accretion rates can result in longer timescales (\ancientTau) for BH being torqued into alignment with the disk.

\item 
\textbf{Gas torques dampen BBH orbital angular momentum}.
Binary black hole systems in AGN disks experience Lindblad and co-rotating gas torques as they migrate through the disk \citep{2020ApJ...898...25T}. 
Co-rotation torques dampen the binary's eccentricity and drive $\vec{L}$ into alignment with $\vec{J}_{\rm AGN}$ on a characteristic timescale \citep{2002ApJ...565.1257T}
\begin{align}
    t_{\rm damp}= \frac{M_{\rm SMBH}^{2} h^{4}}{m_{\rm b} \Sigma a^{2} \Omega}\ ,
\end{align}
where $M_{\rm SMBH}$ is the SMBH mass, $m_{\rm b}$ is the BBH system's total mass, $h=H/a$ the disk aspect ratio, $\Sigma$ is the disk surface density and $\Omega$ is the Keplerian orbital frequency. 
In addition to aligning $\vec{L}$ with $\vec{J}_{\rm AGN}$, dynamical gas friction can promote binary hardening---the process of losing orbital energy and tightening the orbit~\citep{Baruteau:2011:ApJ}.
If $t_{\rm damp}$ is smaller than the lifetime of the AGN (typically in long-lived AGN \oldTau, Table~\ref{tab:astrophysical-imp}ac) the BBH orbital angular momentum vector $\vec{L}$ will align with $\vec{J}_{\rm AGN}$ by the time the binary merges.

\item
\textbf{Dynamical encounters excite BBH orbital angular momentum}.
AGNs' dense environment and high escape velocity facilitate dynamical encounters~\citep{2019ApJ...876..122Y,2019PhRvL.123r1101Y,2020arXiv201009765S}. 
Multiple migrators can quickly interact with each other, potentially leading to complex or chaotic dynamical encounters moderated by the disk gas \citep{2021arXiv211003698W}.
Tertiary encounters of binaries with compact objects in the disk or the spherical nuclear population component can harden or soften BBH systems \citep{2018MNRAS.474.5672L,2019ApJ...876..122Y}, increase BBH orbital eccentricity \citep{2020arXiv201009765S}, and alter the orbital angular momentum of the BBH \citep{2020ApJ...898...25T, Tagawa:2020:ApJ}.
Close encounters with a tertiary object on a disk-crossing orbit can perturb the orbital angular momentum of the BBH on the timescale for the encounter ($t_{\rm enc}$ \citet{2018MNRAS.474.5672L}), which depends on the density of the nuclear star cluster ($\rho_{\rm NSC}$), BBH location in the disk, and the efficiency of disk capture (a function of $\overline{\rho},\overline{\tau}_{\rm AGN}$). Small values of $t_{\rm enc}$ (Table~\ref{tab:astrophysical-imp}cd) lead to more binaries with $\vec{L}$ misaligned with $\vec{J}_{\rm AGN}$ at merger.

\end{enumerate}

The relation between the dampening and dynamical encounter timescales may provide further details about the AGN. For example, \tdampBig\ could occur if 
(I) AGN disks are not long-lived \midTau, and the spherical population is not efficiently captured by the disk, resulting in lots of dynamic interactions~\citep{2018ApJ...866...66M, 2020ApJ...898...25T},
(II) BBH are positioned in a short-lived inner disk where the encounter rate with the spherical component is significantly higher  ($a$ is small, \citet{2018MNRAS.474.5672L}), 
(III) $\rho_{\rm NSC}$ is large (e.g., the NSC is cuspy not cored~\citet{2018ApJ...866...66M,2020ApJ...898...25T}), or 
(IV) The fraction of BBH systems hardened via gas torques $\epsilon_{\rm gas}$\footnote{where $\epsilon_{\rm gas}\to1$ implies all BBH systems have hardened by gas torques, and $\epsilon_{\rm gas}\to0$ implies all BBH systems have hardened by dynamic encounters.} is not particularly efficient relative to dynamical hardening~\citep{2017MNRAS.464..946S}.

In summary, our assumptions are that: 1) BH embedded in AGN disks accrete at the Eddington rate and are thereby gradually torqued into alignment with the disk, 2) BH from the spheroid are captured by the disk over its lifetime (preferentially heavier BH at lower inclination angles), such that heavier BH tend to spend more time embedded in the AGN disk and therefore experience longer periods of torquing. 
Torquing into alignment depends on the gas mass accreted (i.e., a combination of the disk gas density and the disk lifetime for a given accretion rate) and how long a BH has been in the disk. 
For example, a short-lived but dense gas disk can more rapidly torque embedded BH into alignment than a less dense gas disk. 
However, BH that are captured by such a disk will spend less time embedded and will experience a shorter duration torque towards alignment. 

From such assumptions, old (i.e., long-lived), dense AGNs (Table~\ref{tab:astrophysical-imp}a) produce a population of BBH systems with $\vec\chi_1$ preferentially aligned with the orbital angular momentum and $\vec\chi_2$ preferentially aligned with $\vec\chi_1$.
Mid-aged, dense AGNs (Table~\ref{tab:astrophysical-imp}b) produce BBH systems with $\vec\chi_1$ preferentially aligned with the orbital angular momentum, but $\vec\chi_2$ is not correlated with $\vec\chi_1$. This is because we assume $M_{1}$ spends more time embedded in the AGN disk on average. Old, dilute AGNs (Table~\ref{tab:astrophysical-imp}c) produce BBH systems where $\vec\chi_1$ is not preferentially aligned with the orbital angular momentum.
However, $\vec\chi_2$ is preferentially aligned with $\vec\chi_1$.
Young, dilute AGNs (Table~\ref{tab:astrophysical-imp}d) produce BBH systems where $\vec\chi_1$ is uncorrelated with the orbital angular momentum and $\vec\chi_2$ is uncorrelated with $\vec\chi_1$.

\section{Black-hole spin orientation model}

\subsection{Model Description}
\label{sec:model}
\begin{figure}[h!]
\centering 
{\includegraphics[width=0.75\linewidth]{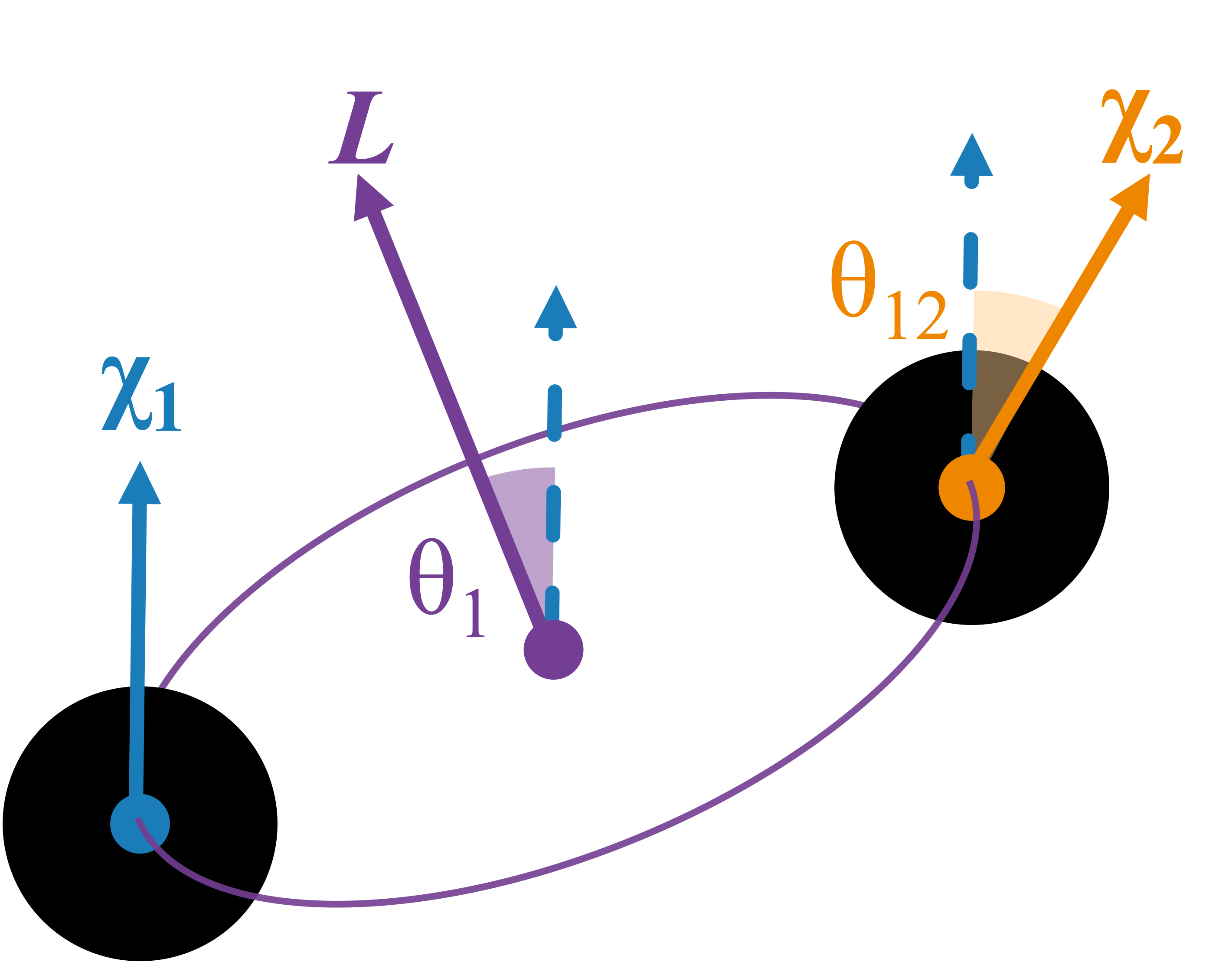}}
\caption{
\textbf{Schematic diagram of a binary black hole system.} 
The blue, orange, and purple arrows correspond to
$\pvec{\chi}_1, \pvec{\chi}_2$ and $\pvec{L}$. The angle between $(\pvec{\chi}_1, \pvec{L})$ is $\theta_1$, while the angle between $(\pvec{\chi}_1, \pvec{\chi_2})$ is $\theta_{12}$.
}
\label{fig:diagram}
\end{figure}

In this section, we construct a model for the spin orientation of black holes (at the time of formation) in merging binaries residing in the AGN disk. 
It will be useful to employ two coordinate systems.
Vectors with no prime are measured with respect to the orbital angular momentum vector such that
\begin{align}
    \hat{z} \propto \vec{L} , 
\end{align}
while primed vectors are measured with respect to the primary spin vector such that
\begin{align}
    \hat{z}' \propto \vec\chi_1 .
\end{align}
Hence, $\vec\chi_1$ is given by
\begin{align}
    \vec\chi_1 = \chi_1 \left(
    \begin{array}{c}
        \sin\theta_1 \cos\phi_1 \\
        \sin\theta_1 \sin\phi_1 \\
        \cos\theta_1
    \end{array}
    \right) ,
\end{align}
in the unprimed coordinate system and
\begin{align}
    \pvec{\chi}'_1 = \chi_1 \left(
    \begin{array}{c}
        0 \\
        0 \\
        1
    \end{array}
    \right) ,
\end{align}
in the primed coordinate system. 
A schematic diagram depicting $\vec\chi_1, \vec\chi_2, \vec L$ and the angles $\theta_1,\theta_{12}$ can be seen in Figure~\ref{fig:diagram}.

Here, $\theta_1$ is the zenith angle between $\vec{L}$ at formation and $\hat\chi_1$, and $\phi_1$ is the azimuthal angle measured from $\hat{x}$, about $\hat{z}'$ at formation.
The secondary spin vector is
\begin{align}
    \pvec{\chi}'_2 = \chi_2 \left(
    \begin{array}{c}
        \sin\theta_{12}\cos\zeta_{12} \\
        \sin\theta_{12}\sin\zeta_{12} \\
        \cos\theta_{12}
    \end{array}
    \right) ,
\end{align}
where $\theta_{12}$ is the zenith angle between the $\vec\chi_1$ and $\vec\chi_2$, and $\zeta_{12}$ is the azimuthal angle measured from $\hat{x}'$, about $\vec\chi_1$.

Our population model is framed in terms of $\theta_1$ and $\theta_{12}$, the at-formation angles between $\vec{L}, \hat\chi_1$  and $\hat\chi_1, \hat\chi_2$ respectively.
The distributions of $\theta_1, \theta_{12}$, denoted with $\pi(...)$, are conditional on hyper-parameters $\sigma_1, \sigma_{12}$, which encodes AGN physics:
\begin{align}
    \pi(\theta_1, \theta_{12} | \sigma_1, \sigma_{12}) .
\end{align}
The hyper-parameters determine the \textit{shape} of the distribution.

Using the parameterisation from~\cite{spin_population_models}, we assume that the cosine of the primary spin zenith angle $\cos\theta_1$ at formation is drawn from a truncated normal distribution (denoted \Nt) with mean=1 and standard deviation $\sigma_1$:
\begin{align}
    \pi(\cos\theta_1 | \sigma_1) = \Nt(\cos\theta_1 | \sigma_1) .
\end{align}
This distribution allows for preferred alignment between $\hat\chi_1$ and $\vec{L}$ with a free parameter $\sigma_1$ controlling the typical misalignment angle.
For small values of $\sigma_1$, the $\hat\chi_1$ distribution tends to be nearly aligned with the $\vec{L}$.
As $\sigma_1$ becomes large, the distribution becomes uniform, so $\hat\chi_1$ is uncorrelated with $\vec{L}$.
We assume the primary azimuthal spin angle $\phi_1$ is drawn from a uniform distribution denoted $U$. 

We assume that the secondary spin vector is preferentially aligned to the primary spin vector
\footnote{We take the phrase `preferentially aligned spin' to mean that the directions of $\pvec{\chi}_1$ and $\pvec{\chi}_2$ vectors are correlated so that they point more nearly in the same direction than two random vectors.}
at formation such that
\begin{align}
    \pi(\cos\theta_{12} | \sigma_{12}) = \Nt(\cos\theta_{12} | \sigma_{12}) .
\end{align}
Small values of $\sigma_{12}$ imply that $\vec\chi_1$ and $\vec\chi_2$ tend to point in nearly the same direction. 
As $\sigma_{12}$ becomes large, the directions of $\vec\chi_1$ and $\vec\chi_2$ become uncorrelated.
We assume that $\zeta_{12}$ is drawn from a uniform distribution. 

\begin{figure}
\centering 
\makebox[\columnwidth]{\includegraphics[width=0.75\linewidth]{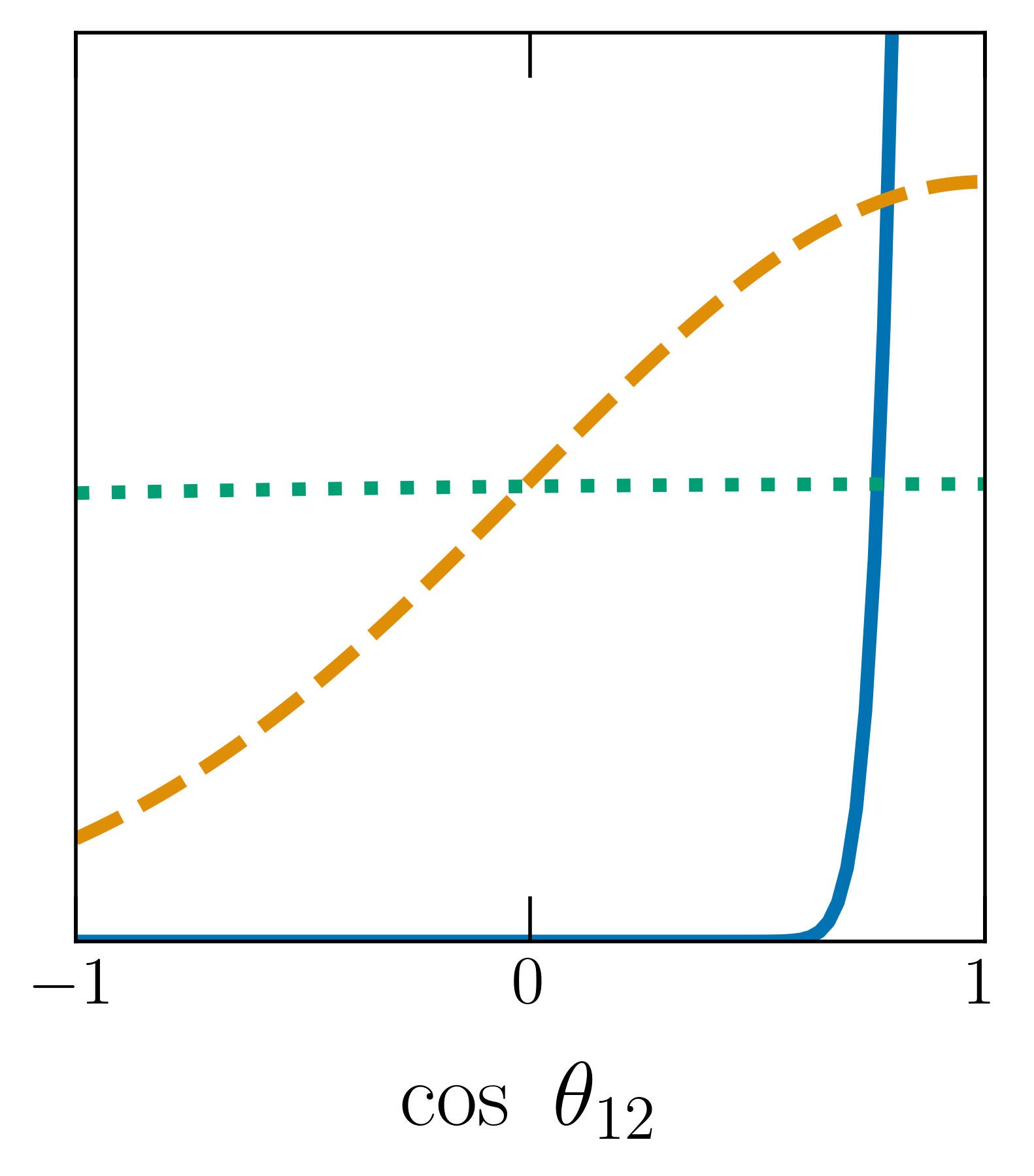}}
\caption{
\textbf{\boldmath The distribution of $\cos\theta_{12}$ for two different values of $\sigma_{12}$.\unboldmath} 
The blue solid curve, orange dashed curve, and green dotted curve correspond to $\sigma_{12}=10^{-1}, 10^{0}, 10^{1}$ respectively. 
Small values of $\sigma_{12}$ model relatively old AGNs, while larger values model relatively younger AGNs.
As $\sigma_{12} \to 0$ (the blue solid curve), $\vec\chi_1$ and $\vec\chi_2$ become aligned. 
On the other hand, when $\sigma_{12}\to\infty$ (the green dotted curve), $\vec\chi_2$ becomes uncorrelated with $\vec\chi_1$.
}\label{fig:cos_theta12_examples}
\end{figure}

Putting everything together, the prior for the spin vector orientations at formation is given by 
\begin{align}\label{eq:agn_model}
    \pi(\theta_1, \theta_{12} | \sigma_1, \sigma_{12}) = \ 
    &\Nt(\cos\theta_1 | \sigma_1)\ 
    \Nt(\cos\theta_{12} | \sigma_{12})\nonumber \\
    &U(\phi_1) \ U(\zeta_{12}) ,
\end{align}
where $U(\phi_1), U(\zeta_{12})$ are constants equal to $1/2\pi$.
Figure~\ref{fig:cos_theta12_examples} displays distributions of $\cos\theta_{12}$ for different values of $\sigma_{12}$. 
In the following section, we delve into various configurations of $\sigma_1$ and $\sigma_{12}$ and their physical implications on AGNs. 

\subsection{The evolution of spin vectors with time}\label{evolution}

\begin{figure}
    \centering
    \includegraphics[width=0.85\linewidth]{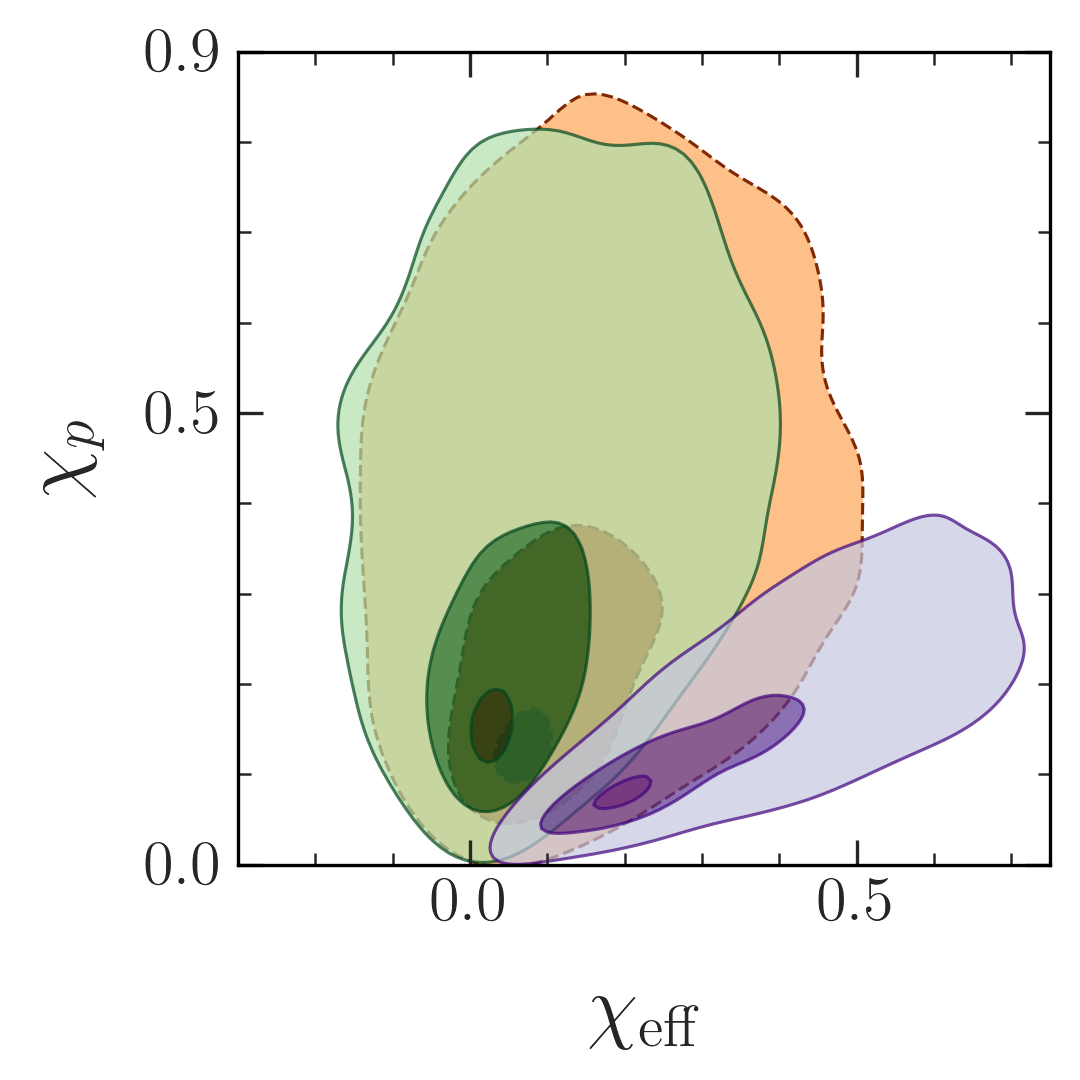}
    \caption{
    \textbf{\boldmath 
    Plots of $\pi(\chi_p, \chi_{\rm eff} | \sigma_1, \sigma_{12})$ distributions for three binary-black hole spin populations.
    \unboldmath}
    The hyper-parameters for each distribution are:
    dashed-orange $\sigma_1=0.1, \sigma_{12}=10$,
    solid-purple $\sigma_1=0.1, \sigma_{12}=0.1$, and
    solid-green $\sigma_1=10, \sigma_{12}=0.1$.
    Contours are drawn at the 1,2,3-sigma levels.
    \label{fig:xeffxp2d}}
\end{figure}

As BBH systems evolve, coupling of the component black hole spin vectors and the orbital angular momentum vector results in Lense-Thirring precession, causing the spin and angular momenta to precess about the system's total angular momentum vector~\citep{Mashhoon:1984:GReGr}.
Thus, the angles $\theta_1, \theta_{12}$ evolve over time.
While the distributions of $\theta_1, \theta_{12}$ modeled in Eq.~\ref{eq:agn_model} describe the binary properties at formation (approximately when the binary is infinitely separated, \citet{Johnson-McDaniel:2021:arXiv}), these distributions are different by the time these binaries enter the observing band of audio-band gravitational-wave detectors (20-20,000 Hz).

Fortunately, the information present in the distribution of $\theta_1, \theta_{12}$ at formation is encoded in the distribution of the ``effective inspiral spin parameter'' $\chi_{\rm eff}$ \citep{Damour2001} and the effective precession parameter $\chi_p$ \citep{Schmidt2012}, which are approximate constants of motion~\citep{Hannam:2014:PhRvL, Gerosa:2021:PhRvD}. 
The parameter $\chi_{\rm eff}$ measures the spin components aligned with the orbital angular momentum while $\chi_p$ measures the spin components in the orbital plane.

Figure~\ref{fig:xeffxp2d} shows joint distributions of $\chi_{\rm eff} ,\chi_p$, each representing three different AGN populations, each with different values of $\sigma_1, \sigma_{12}$.
The dashed-orange distribution is created using $\sigma_1=0.1$ and $\sigma_{12} = 10$ (a mid-aged, dense AGN). 
The solid-purple distribution displays the distribution for $\sigma_1=0.1$ and $\sigma_{12} = 0.1$ (an old, dense AGN).
Finally, the solid-green contours is the distribution with $\sigma_1=10$ and $\sigma_{12} = 10$ (a young, dilute AGN).
The distinguishability of these distributions illustrates how the AGN properties may be imprinted on the distribution of quantities measured by LIGO--Virgo.

To recast our model in terms of quantities that are measured by LIGO--Virgo, it is necessary to compute
\begin{align}\label{eq:eff_prior}
    \pi(\chi_\text{eff}, \chi_p | \sigma_1, \sigma_{12}) .
\end{align}
In principle, an expression for this distribution may be obtained through a series of convolutional integrals, which most likely have to be evaluated numerically.
An alternative to numerical integration is estimating the probability density distribution at fixed values of $\sigma_1, \sigma_{12}$ using histograms or kernel density estimators.
These estimates can be used to interpolate the probability density for arbitrary values of $\sigma_1, \sigma_{12}$, for example, using a machine learning algorithm; see \citet{HernandezVivanco:2020:MNRAS}.
Alternatively, a machine learning algorithm can likely be trained to reproduce the results of numerical integrals for different values of $\sigma_1, \sigma_{12}$.

However, as this work focuses on estimating how well LIGO--Virgo will be able to measure $\sigma_1, \sigma_{12}$, creating a machine-learning representation of Eq.~\ref{eq:eff_prior} goes beyond our present scope. 
Instead, in the demonstration that follows in Section~\ref{sec:results}, we pretend that the values of $\theta_1, \theta_{12}$ measured by LIGO--Virgo are unchanged since the binary was formed.
This is tantamount to assuming that the $\chi_\text{eff}, \chi_p$ distribution encodes---without loss---all of the information in the distribution of $\theta_1, \theta_{12}$.
Since there is likely information loss, our results are overly optimistic.
Nonetheless, judging by the distinguishability of different populations shown in Fig.~\ref{fig:xeffxp2d}, we believe this assumption still yields a ballpark estimate.

In Table~\ref{tab:astrophysical-imp}, we summarize the implications of measuring various $\sigma_1$ and $\sigma_{12}$ on some parameters describing AGNs. 
Some of the implications in Tab.~\ref{tab:astrophysical-imp} are degenerate.
For example, if AGN disks are relatively low density on average (\lowRho), then their corresponding average lifetimes \oldTau\ to torque $\vec{\chi}_{1,2}$ into alignment with $\vec{J}_{\rm AGN}$. 
Moreover, the top left quadrant (small $\sigma_{1},\sigma_{12} \leq 1$, Table~\ref{tab:astrophysical-imp}a) is indistinguishable from a population of BBH systems assembled in the field binary (aligned spins).
The bottom right quadrant (large $\sigma_{1},\sigma_{12} > 1$, Table~\ref{tab:astrophysical-imp}d) is consistent with a dynamical assembly in a dense stellar environment such as a globular cluster (isotropic spins). 
Thus, depending on the nature of BBH assembly in AGN disks, this framework does not necessarily provide a useful means of testing the AGN hypothesis against competing hypotheses.
Rather, it is a means of probing AGN physics \textit{assuming} the AGN hypothesis is true.
Independent evidence, e.g., from electromagnetic counterparts, may be required to establish this premise.

\section{Demonstration}\label{sec:results}
In this section, we carry out a demonstration analysis to estimate the ability of Advanced LIGO--Virgo to measure AGN physics.
We generate two simulated populations of one hundred BBH signals drawn from our population model described by Eq.~\ref{eq:agn_model}. 
The BBH systems are uniformly distributed in comoving volume. 
We employ the mass model from \cite{mass_population_models} with parameters consistent with results from \cite{Abbott:2021:ApJL}.
For the sake of simplicity, we assume fixed values for the mass-ratio $q=1$ and dimensionless spin magnitude $\chi_1=0.6, \chi_2=0.6$.
These simplifying assumptions roughly match the mass-ratio and spin magnitudes of the 20\% of merging BBH systems with non-zero spin \citep{Roulet,BuildingBetterModels}. 
We additionally use an arbitrary polarisation angle $\phi=0.1$ to simplify analysis further.
We draw the black hole spin tilts for each population from:
\begin{align*}
    \blacktriangleright\, \text{Population A: } &\pi_{\rm{AGN}}(\vec\chi_1, \vec\chi_2 \, |\ \sigma_1=0.5,\  \sigma_{12}=1),\\
    \blacktriangleright\,\text{Population B: } &\pi_{\rm{AGN}}(\vec\chi_1, \vec\chi_2 \, |\ \sigma_1=3.0,\ \sigma_{12}=0.25)
\end{align*}
at a reference frequency of $20\ \text{Hz}$\footnote{We use 20 Hz for consistency with GWTC-2~\citep{gwtc-2}, a frequency close to the lower-end of the LIGO sensitivity curve~\citep{ligo, abbott_19_observing_scenarios}. Note that a reference frequency will not be required when switching to $\pi(\chi_\text{eff}, \chi_p | \sigma_1, \sigma_{12})$ from $\pi(\cos \theta_{1}, \cos \theta_{12} | \sigma_1, \sigma_{12})$.}. 
As discussed in Section~\ref{evolution}, we ignore the evolution of the spin vectors from the time of formation to the moment they enter the LIGO--Virgo band, which is equivalent to assuming that the information in the $\theta_1, \theta_{12}$ distribution is encoded in the distribution of $\chi_\text{eff}, \chi_p$ without any data loss.
This assumption makes our results overly optimistic.
Population A corresponds to mergers from AGNs that are old and dense (Table~\ref{tab:astrophysical-imp}a). 
On the other hand, Population B corresponds to mergers from AGNs that are old and dilute (Table~\ref{tab:astrophysical-imp}c). 

\begin{table*}
\centering
\caption{
\textbf{Population and Prior distributions for demonstration.}
We use shorthand to represent distributions: $\delta$~delta, $U$~uniform, $CM$~uniform in comoving volume, $C$~cosine, and finally $\mathcal{N}_t$~truncated normal (mean of 1).
The definitions of the parameters are documented in \citet[Table~E1]{Romero-Shaw:2020:MNRAS}.
The `Population' and `Prior'  columns display the spin orientation distributions for the population and prior. The `Common' displays some parameter distributions common between the population and prior.
}
\label{tab:pop_and_pri_dist}
\begin{tabular}{rllrllrll}
\multicolumn{2}{c}{Population}       &  & \multicolumn{2}{c}{Prior}             &  & \multicolumn{2}{c}{Common} &  \\
Parameter             & Distribution &  & Parameter              & Distribution &  & Parameter   & Distribution &  \\ \cline{1-2} \cline{4-5} \cline{7-8}
$|\vec{\chi}_1|, |\vec{\chi}_2|$ &
  $\delta(0.6)$ &
   &
  $|\vec{\chi}_1|, |\vec{\chi}_2|$ &
  $\delta(0.6)$ &
   &
  $\mathcal{M}/M_{\odot}$ &
  $U(15,60)$ &
   \\
$\cos\theta_1$ &
  $\mathcal{N}_t(\sigma_1)$ &
   &
  $\cos\theta_1$ &
  $U(-1,1)$ &
   &
  $q$, $\phi$ &
  $\delta(1),  \delta(0.1)$ &
   \\
$\cos\theta_{12}$ &
  $\mathcal{N}_t(\sigma_{12})$ &
   &
  $\cos\theta_2$ &
  $U(-1,1)$ &
   &
  $d_{\rm{L}}/\rm{Mpc}$ &
  $CM(200, 800)$ &
   \\
$\eta_{12}, \phi_{1}$ & $U(0,2\pi)$  &  & $\phi_{12}, \phi_{JL}$ & $U(0,2\pi)$  &  & dec         & $C(0, 2\pi)$ &  \\
$\iota$               & $U(0, \pi)$  &  & $\theta_{JN}$          & $S(0, 2\pi)$ &  & ra,  $\psi$ & $U(0, 2\pi)$ &  \\ 
\end{tabular}
\end{table*}

We simulate the gravitational-wave signals from the BBH mergers using the waveform approximant \imrxhm~\citep{Pratten:2020fqn, Pratten:2020ceb, Garcia-Quiros:2020qpx}. 
We add the simulated signals into Gaussian noise coloured to the Advanced LIGO design sensitivity for the Hanford and Livingston detectors~\citep{ligo, abbott_19_observing_scenarios}. 
We ensure that the matched-filter signal-to-noise ratio $\text{SNR}\geq8$. 
We perform Bayesian inference with \parallelbilby \citep{bilby_paper, pbilby_paper,Romero-Shaw:2020:MNRAS, dynesty_paper, skilling2004, skilling2006} and \gwpop \citep{gwpopulation} to recover posterior probability densities for the parameters of the simulated signals using the same waveform approximant \imrxhm, also evaluated at a reference frequency of $20\ \text{Hz}$. 
The prior and population probability distributions we use are documented in Table~\ref{tab:pop_and_pri_dist}. 

\begin{figure}
\centering 
\makebox[\columnwidth]{\includegraphics[width=0.98\linewidth]{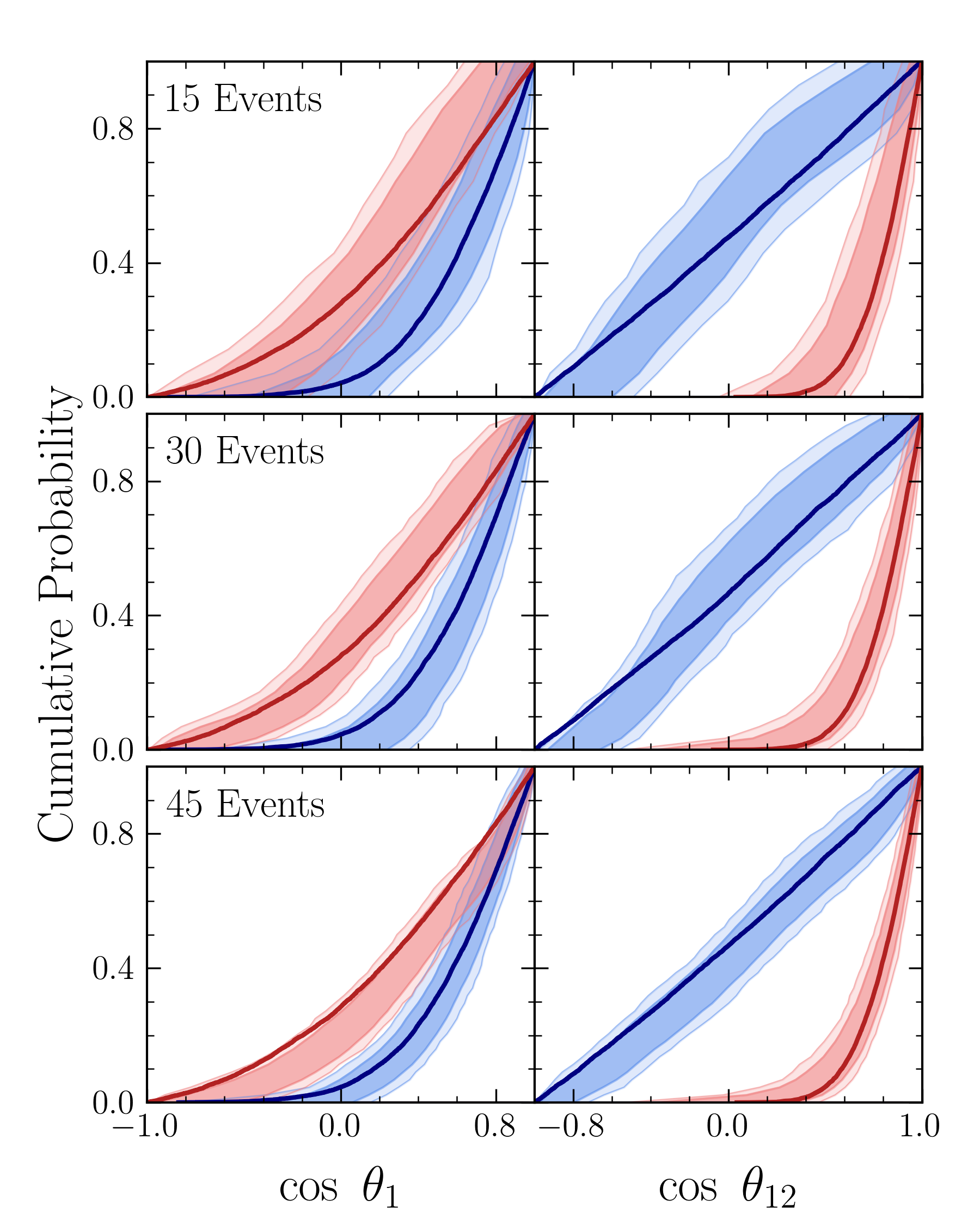}}
\caption{
\textbf{
\boldmath
Cumulative distributions of $\cos\theta_1$ and $\cos\theta_{12}$. 
\unboldmath
} 
Population A is shaded blue while Population B is shaded red. 
The bands show the $99\%$ (light shading) and $90\%$ (darker shading)  credible intervals of the observed posterior distributions of the simulated catalogs. 
The solid curve (darkest) shows the true population distribution. 
 }
\label{fig:result_cdfs}
\end{figure}

In Fig.~\ref{fig:result_cdfs} we plot cumulative distributions of $\cos\theta_1, \cos\theta_{12}$ for Population A (old and dense, blue) and Population B (old and dilute, red). 
The true population distributions are the solid dark curves. 
The empirically observed distributions from the simulated catalogs are the shaded bands. 
The darker region of the shaded bands indicates the $90\%$ credibility range of the observed distributions, while the lighter region indicates the $99\%$ credibility range. 
Each row of the figure displays the observed distributions for different sizes of the populations. 
As the number of events in each population increases, the observed $99\%$ credibility ranges shrink. 
Figure~\ref{fig:result_cdfs} demonstrates that the two populations can be visually distinguished once there are $\mathcal{O}(10)$ gravitational-wave events.
However, given recent results that suggest only $\approx20\%$ of BBH systems contain a black hole with measurable spin \citep{Roulet,BuildingBetterModels,Miller2020}---and given the loss of information moving from $\theta_1,\theta_{12}$ to $\chi_\text{eff}, \chi_p$---it likely that $\gtrsim\mathcal{O}(50)$ events are required.

\section{Discussion}
This paper introduces a physically motivated phenomenological population model describing the at-formation spin-orientations of merging binary black holes assembled in active galactic nuclei. 
By measuring the distribution of spin orientations with gravitational waves, we may be able to learn about the AGN environment; whether it is old or young, whether it is dilute or dense.
We demonstrate that $\mathcal{O}(10)$ gravitational-wave events from BBH mergers with $\text{SNR}\geq8$ and spin magnitudes $\chi\sim0.6$ are required to infer the population parameters describing the shape of the $\cos\theta_1$ and $\cos\theta_{12}$ distributions.
However, since the majority of BBH mergers include black holes with negligible spin \citep{Roulet,BuildingBetterModels,Miller2020}---and since we have optimistically assumed that the distribution of $\chi_\text{eff}, \chi_p$ retain all the information of the spin orientations at formation---a total of $\gtrsim\mathcal{O}(50)$ BBH detections are necessary to begin to resolve the properties of AGN. 

To derive this estimate, we assume that all BBH mergers take place in AGNs.
This is a reasonable starting point since it is desirable to see if all BBH detections can be understood within a single channel, and it would be somewhat surprising if it turns out that two different channels produce comparable merger rates.
However, it may turn out that AGNs provide only some fraction of the observed population.
If so, our effort to infer the properties of AGN disks will be complicated by contamination from other channels.
One could use the model proposed here as a sub-population in a mixture model, but this goes beyond our present scope.

Our model is cast in terms of the orientation of spin vectors at the time of BBH formation.
While the spin vectors subsequently evolve due to precession, the initial orientation of spin vectors is imprinted on the effective spin parameters $\chi_\text{eff}, \chi_p$; see Fig~\ref{fig:xeffxp2d}.
To further develop this model such that it can be used for Bayesian inference, it will be necessary to recast the model in terms of these effective parameters (see Eq.~\ref{eq:eff_prior}).
Future work will focus on developing a computationally efficient representation of this distribution, for example, using a machine learning algorithm.
With a computationally efficient model, it will be possible to apply the model to current gravitational-wave catalogs, which may now have enough events to begin resolving properties of AGN physics.

\section*{Acknowledgments}{

We thank Shanika Galaudage and Colm Talbot for their technical help. 
We thank Ilya Mandel for their helpful advice.

We gratefully acknowledge the Swinburne Supercomputing OzSTAR Facility for computational resources. All analyses (including test and failed analyses) performed for this study used $130$K core-hours on OzSTAR. This would have amounted to a carbon footprint of ${\sim \unit[7.6]{\text{t CO}_2}}$~\citep{greenhouse}. However, as OzSTAR is powered by wind energy from Iberdrola Australia, the electricity for computations produces negligible carbon waste.

This material is based upon work supported by NSF’s LIGO Laboratory, a major facility fully funded by the National Science Foundation. This research has used data, software, and web tools obtained from the Gravitational Wave Open Science Center (\href{https://www.gw-openscience.org}{https://www.gw-openscience.org}), a service of LIGO Laboratory, the LIGO Scientific Collaboration, and the Virgo Collaboration. LIGO Laboratory and Advanced LIGO are funded by the United States National Science Foundation (NSF) as well as the Science and Technology Facilities Council (STFC) of the United Kingdom, the Max-Planck-Society (MPS), and the State of Niedersachsen/Germany for support of the construction of Advanced LIGO and construction and operation of the GEO600 detector. The Australian Research Council provided additional support for Advanced LIGO. Virgo is funded, through the European Gravitational Observatory (EGO), the French Centre National de Recherche Scientifique (CNRS), the Italian Istituto Nazionale di Fisica Nucleare (INFN) and the Dutch Nikhef, with contributions by institutions from Belgium, Germany, Greece, Hungary, Ireland, Japan, Monaco, Poland, Portugal, Spain.

AV, RS \& ET are supported by the Australian Research Council (ARC) Centre of Excellence CE170100004. BM \& KESF are supported by NSF AST-1831415 and Simons Foundation Grant 533845.
}

\vspace{5mm}

\software{
\href{https://lscsoft.docs.ligo.org/parallel_bilby/}{\parallelbilby}~\citep[v1.0.1]{parallel_bilby},
\href{https://lscsoft.docs.ligo.org/bilby/}{\bilby}~\citep[v1.1.4]{bilby},
\href{https://lscsoft.docs.ligo.org/bilby_pipe/master/index.html}{\bilbypipe}~\citep[v1.0.4]{bilby_pipe},
\href{https://pypi.org/project/gwpopulation/}{\gwpop}~\citep[v0.6.3]{gwpopulation},
\href{https://dynesty.readthedocs.io/}{\dynesty}~\citep[v1.0.1]{dynesty_paper},
\href{https://gwpy.github.io/docs/stable/index.html}{\gwpy}~\citep[v2.1.0]{gwpy},
\href{https://lscsoft.docs.ligo.org/lalsuite/lalsimulation/index.html}{\code{LALSimulation}}~\citep[v7.0]{lalsuite},
\code{matplotlib}~\citep[v3.2.0]{matplotlib}, \code{NumPy}~\citep[v1.8.1]{NumPy},
\code{SciPy}~\citep[v1.4.1]{SciPy},
\code{pandas}~\citep[v1.0.2]{pandas},
\code{python}~\citep[v3.7]{pythonForScientificComputing,pythonForScientists}. 
}

\bibliographystyle{aasjournal}
\bibliography{main}

\begin{thebibliography}{}
\expandafter\ifx\csname natexlab\endcsname\relax\def\natexlab#1{#1}\fi
\providecommand{\url}[1]{\href{#1}{#1}}

\bibitem[{{Abbott} {et~al.}(2018){Abbott}, {Abbott}, {Abbott},
  {et~al.}}]{abbott_19_observing_scenarios}
{Abbott}, B.~P., {Abbott}, R., {Abbott}, T.~D., {et~al.} 2018, Living Reviews
  in Relativity, 21, 3

\bibitem[{Abbott {et~al.}(2019)}]{o2_pop}
Abbott, B.~P., {et~al.} 2019, Astrophys. J. Lett., 882, L24

\bibitem[{{Abbott} {et~al.}(2021){Abbott}, {Abbott}, {Abraham}, {Acernese}, \&
  et~al.}]{Abbott:2021:ApJL}
{Abbott}, R., {Abbott}, T.~D., {Abraham}, S., {Acernese}, F., \& et~al. 2021,
  \apjl, 913, L7

\bibitem[{{Acernese} {et~al.}(2015)}]{virgo}
{Acernese}, F., {et~al.} 2015, Classical Quantum Gravity, 32, 024001

\bibitem[{{Antonini} \& {Rasio}(2016)}]{2016ApJ...831..187A}
{Antonini}, F., \& {Rasio}, F.~A. 2016, \apj, 831, 187

\bibitem[{{Ashton} {et~al.}(2019{\natexlab{a}}){Ashton}, {H{\"u}bner}, {Lasky},
  \& {Talbot}}]{bilby}
{Ashton}, G., {H{\"u}bner}, M., {Lasky}, P., \& {Talbot}, C.
  2019{\natexlab{a}}, {Bilby: A User-Friendly Bayesian Inference Library},
  v0.4.0,  Zenodo, doi:10.5281/zenodo.2602178

\bibitem[{Ashton {et~al.}(2020)Ashton, Romero-Shaw, Talbot, Hoy, \&
  Galaudage}]{bilby_pipe}
Ashton, G., Romero-Shaw, I., Talbot, C., Hoy, C., \& Galaudage, S. 2020, {bilby
  pipe: 1.0.1}, vv1.0.1,  https://git.ligo.org.
\newblock \url{https://lscsoft.docs.ligo.org/bilby_pipe/master/index.html}

\bibitem[{{Ashton} {et~al.}(2019{\natexlab{b}}){Ashton}, {H{\"u}bner}, {Lasky},
  {Talbot}, {Ackley}, {Biscoveanu}, {Chu}, {Divakarla}, {Easter}, {Goncharov},
  {Hernandez Vivanco}, {Harms}, {Lower}, {Meadors}, {Melchor}, {Payne},
  {Pitkin}, {Powell}, {Sarin}, {Smith}, \& {Thrane}}]{bilby_paper}
{Ashton}, G., {H{\"u}bner}, M., {Lasky}, P.~D., {et~al.} 2019{\natexlab{b}},
  \apjs, 241, 27

\bibitem[{{Australian Government - Department of the Environment and
  Energy}(2021)}]{greenhouse}
{Australian Government - Department of the Environment and Energy}. 2021,
  {National Greenhouse Gas Inventory: Quarterly updates},  {Department of the
  Environment and Energy}.
\newblock
  \url{https://www.industry.gov.au/data-and-publications/national-greenhouse-gas-inventory-quarterly-updates}

\bibitem[{{Bartos} {et~al.}(2017){Bartos}, {Kocsis}, {Haiman}, \&
  {M{\'a}rka}}]{2017ApJ...835..165B}
{Bartos}, I., {Kocsis}, B., {Haiman}, Z., \& {M{\'a}rka}, S. 2017, \apj, 835,
  165

\bibitem[{{Baruteau} {et~al.}(2011){Baruteau}, {Cuadra}, \&
  {Lin}}]{Baruteau:2011:ApJ}
{Baruteau}, C., {Cuadra}, J., \& {Lin}, D.~N.~C. 2011, \apj, 726, 28

\bibitem[{{Bogdanovi{\'c}} {et~al.}(2007){Bogdanovi{\'c}}, {Reynolds}, \&
  {Miller}}]{2007ApJ...661L.147B}
{Bogdanovi{\'c}}, T., {Reynolds}, C.~S., \& {Miller}, M.~C. 2007, \apjl, 661,
  L147.
\newblock \url{https://arxiv.org/pdf/astro-ph/0703054.pdf}

\bibitem[{Damour(2001)}]{Damour2001}
Damour, T. 2001, Phys. Rev. D, 64, 124013

\bibitem[{{Dominik} {et~al.}(2013){Dominik}, {Belczynski}, {Fryer}, {Holz},
  {Berti}, {Bulik}, {Mandel}, \& {O'Shaughnessy}}]{2013ApJ...779...72D}
{Dominik}, M., {Belczynski}, K., {Fryer}, C.~i., {et~al.} 2013, Astrophys. J.,
  779, 72

\bibitem[{{Eldridge} {et~al.}(2017){Eldridge}, {Stanway}, {Xiao}, {McClelland},
  {Taylor}, {Ng}, {Greis}, \& {Bray}}]{2017PASA...34...58E}
{Eldridge}, J.~J., {Stanway}, E.~R., {Xiao}, L., {et~al.} 2017, Pub. Astron.
  Soc. Aust., 34, e058

\bibitem[{{Fabj} {et~al.}(2020){Fabj}, {Nasim}, {Caban}, {Ford}, {McKernan}, \&
  {Bellovary}}]{2020MNRAS.499.2608F}
{Fabj}, G., {Nasim}, S.~S., {Caban}, F., {et~al.} 2020, \mnras, 499, 2608

\bibitem[{Fishbach {et~al.}(2017)Fishbach, Holz, \& Farr}]{Fishbach:2017dwv}
Fishbach, M., Holz, D.~E., \& Farr, B. 2017, Astrophys. J. Lett., 840, L24

\bibitem[{{Fragione} {et~al.}(2019){Fragione}, {Grishin}, {Leigh}, {Perets}, \&
  {Perna}}]{2019MNRAS.488...47F}
{Fragione}, G., {Grishin}, E., {Leigh}, N. W.~C., {Perets}, H.~B., \& {Perna},
  R. 2019, \mnras, 488, 47

\bibitem[{Galaudage {et~al.}(2021)Galaudage, Talbot, Nagar, Jain, Thrane, \&
  Mandel}]{BuildingBetterModels}
Galaudage, S., Talbot, C., Nagar, T., {et~al.} 2021, Accepted for publication
  in Astrophys. J. Lett., arxiv/2109.02424

\bibitem[{Garc\'\i{}a-Quir\'os {et~al.}(2020)Garc\'\i{}a-Quir\'os, Colleoni,
  Husa, Estell\'es, Pratten, Ramos-Buades, Mateu-Lucena, \&
  Jaume}]{Garcia-Quiros:2020qpx}
Garc\'\i{}a-Quir\'os, C., Colleoni, M., Husa, S., {et~al.} 2020, Phys. Rev. D,
  102, 064002

\bibitem[{{Gayathri} {et~al.}(2020){Gayathri}, {Healy}, {Lange}, {O'Brien},
  {Szczepanczyk}, {Bartos}, {Campanelli}, {Klimenko}, {Lousto}, \&
  {O'Shaughnessy}}]{Gayathri2020}
{Gayathri}, V., {Healy}, J., {Lange}, J., {et~al.} 2020, arXiv e-prints,
  arXiv:2009.05461

\bibitem[{{Generozov} {et~al.}(2018){Generozov}, {Stone}, {Metzger}, \&
  {Ostriker}}]{2018MNRAS.478.4030G}
{Generozov}, A., {Stone}, N.~C., {Metzger}, B.~D., \& {Ostriker}, J.~P. 2018,
  \mnras, 478, 4030

\bibitem[{{Gerosa} {et~al.}(2021){Gerosa}, {Mould}, {Gangardt}, {Schmidt},
  {Pratten}, \& {Thomas}}]{Gerosa:2021:PhRvD}
{Gerosa}, D., {Mould}, M., {Gangardt}, D., {et~al.} 2021, \prd, 103, 064067

\bibitem[{Giacobbo {et~al.}(2017)Giacobbo, Mapelli, \&
  Spera}]{2018MNRAS.474.2959G}
Giacobbo, N., Mapelli, M., \& Spera, M. 2017, MNRAS, 474, 2959

\bibitem[{{Gr{\"o}bner} {et~al.}(2020){Gr{\"o}bner}, {Ishibashi}, {Tiwari},
  {Haney}, \& {Jetzer}}]{2020A&A...638A.119G}
{Gr{\"o}bner}, M., {Ishibashi}, W., {Tiwari}, S., {Haney}, M., \& {Jetzer}, P.
  2020, \aap, 638, A119

\bibitem[{{Hailey} {et~al.}(2018){Hailey}, {Mori}, {Bauer}, {Berkowitz},
  {Hong}, \& {Hord}}]{2018Natur.556...70H}
{Hailey}, C.~J., {Mori}, K., {Bauer}, F.~E., {et~al.} 2018, \nat, 556, 70

\bibitem[{{Hankla} {et~al.}(2020){Hankla}, {Jiang}, \&
  {Armitage}}]{Hankla:2020:ApJ}
{Hankla}, A.~M., {Jiang}, Y.-F., \& {Armitage}, P.~J. 2020, \apj, 902, 50

\bibitem[{{Hannam} {et~al.}(2014){Hannam}, {Schmidt}, {Boh{\'e}}, {Haegel},
  {Husa}, {Ohme}, {Pratten}, \& {P{\"u}rrer}}]{Hannam:2014:PhRvL}
{Hannam}, M., {Schmidt}, P., {Boh{\'e}}, A., {et~al.} 2014, \prl, 113, 151101

\bibitem[{Harris {et~al.}(2020)Harris, Millman, van~der Walt, Gommers,
  Virtanen, Cournapeau, Wieser, Taylor, Berg, Smith, Kern, Picus, Hoyer, van
  Kerkwijk, Brett, Haldane, Fernández~del Río, Wiebe, Peterson,
  Gérard-Marchant, Sheppard, Reddy, Weckesser, Abbasi, Gohlke, \&
  Oliphant}]{NumPy}
Harris, C.~R., Millman, K.~J., van~der Walt, S.~J., {et~al.} 2020, Nature, 585,
  357–362

\bibitem[{{Hernandez Vivanco} {et~al.}(2020){Hernandez Vivanco}, {Smith},
  {Thrane}, \& {Lasky}}]{HernandezVivanco:2020:MNRAS}
{Hernandez Vivanco}, F., {Smith}, R., {Thrane}, E., \& {Lasky}, P.~D. 2020,
  \mnras, 499, 5972

\bibitem[{Hunter(2007)}]{matplotlib}
Hunter, J.~D. 2007, Computing in science \& engineering, 9, 90

\bibitem[{{Johnson-McDaniel} {et~al.}(2021){Johnson-McDaniel}, {Kulkarni}, \&
  {Gupta}}]{Johnson-McDaniel:2021:arXiv}
{Johnson-McDaniel}, N.~K., {Kulkarni}, S., \& {Gupta}, A. 2021, arXiv e-prints,
  arXiv:2107.11902

\bibitem[{{Kalogera}(2000)}]{2000ApJ...541..319K}
{Kalogera}, V. 2000, Astrophys. J., 541, 319

\bibitem[{{Leigh} {et~al.}(2018){Leigh}, {Geller}, {McKernan}, {Ford}, {Mac
  Low}, {Bellovary}, {Haiman}, {Lyra}, {Samsing}, {O'Dowd}, {Kocsis}, \&
  {Endlich}}]{2018MNRAS.474.5672L}
{Leigh}, N.~W.~C., {Geller}, A.~M., {McKernan}, B., {et~al.} 2018, \mnras, 474,
  5672

\bibitem[{{LIGO Scientific Collaboration}(2020)}]{lalsuite}
{LIGO Scientific Collaboration}. 2020, {LALSuite: LIGO Scientific Collaboration
  Algorithm Library Suite},  https://git.ligo.org, ascl:2012.021

\bibitem[{{LIGO Scientific Collaboration} {et~al.}(2015){LIGO Scientific
  Collaboration}, {Aasi}, {Abbott}, {et~al.}}]{ligo}
{LIGO Scientific Collaboration}, {Aasi}, J., {Abbott}, B.~P., {et~al.} 2015,
  Classical and Quantum Gravity, 32, 074001

\bibitem[{{LIGO Scientific Collaboration} {et~al.}(2021){LIGO Scientific
  Collaboration}, {the Virgo Collaboration}, {Abbott}, {Abbott}, \&
  et~al.}]{gwtc-2}
{LIGO Scientific Collaboration}, {the Virgo Collaboration}, {Abbott}, R.,
  {Abbott}, T.~D., \& et~al. 2021, arXiv e-prints, arXiv:2108.01045

\bibitem[{{Liu} \& {Lai}(2017)}]{Liu:2017:ApJL}
{Liu}, B., \& {Lai}, D. 2017, \apjl, 846, L11

\bibitem[{Lower {et~al.}(2018)Lower, Thrane, Lasky, \& Smith}]{eccentricity}
Lower, M.~E., Thrane, E., Lasky, P.~D., \& Smith, R. J.~E. 2018, Phys. Rev. D,
  98, 083028

\bibitem[{Macleod {et~al.}(2020)Macleod, Urban, Coughlin, Massinger, Pitkin,
  paulaltin, Areeda, Quintero, Badger, Singer, \& Leinweber}]{gwpy}
Macleod, D., Urban, A.~L., Coughlin, S., {et~al.} 2020, {gwpy/gwpy: 1.0.1},
  vv1.0.1,  Zenodo, doi:10.5281/zenodo.3598469.
\newblock \url{https://doi.org/10.5281/zenodo.3598469}

\bibitem[{{Mandel} \& {O'Shaughnessy}(2010)}]{2010CQGra..27k4007M}
{Mandel}, I., \& {O'Shaughnessy}, R. 2010, Class. Quant. Grav., 27, 114007

\bibitem[{{Mapelli} \& {Gualandris}(2016)}]{2016LNP...905..205M}
{Mapelli}, M., \& {Gualandris}, A. 2016, {Star Formation and Dynamics in the
  Galactic Centre}, Vol. 905 (Springer), 205

\bibitem[{{Mashhoon} {et~al.}(1984){Mashhoon}, {Hehl}, \&
  {Theiss}}]{Mashhoon:1984:GReGr}
{Mashhoon}, B., {Hehl}, F.~W., \& {Theiss}, D.~S. 1984, General Relativity and
  Gravitation, 16, 711

\bibitem[{{McKernan} {et~al.}(2014){McKernan}, {Ford}, {Kocsis}, {Lyra}, \&
  {Winter}}]{2014MNRAS.441..900M}
{McKernan}, B., {Ford}, K.~E.~S., {Kocsis}, B., {Lyra}, W., \& {Winter}, L.~M.
  2014, \mnras, 441, 900

\bibitem[{{McKernan} {et~al.}(2012){McKernan}, {Ford}, {Lyra}, \&
  {Perets}}]{2012MNRAS.425..460M}
{McKernan}, B., {Ford}, K.~E.~S., {Lyra}, W., \& {Perets}, H.~B. 2012, \mnras,
  425, 460

\bibitem[{{McKernan} {et~al.}(2020){McKernan}, {Ford}, \&
  {O'Shaughnessy}}]{2020MNRAS.498.4088M}
{McKernan}, B., {Ford}, K.~E.~S., \& {O'Shaughnessy}, R. 2020, \mnras, 498,
  4088

\bibitem[{{McKernan} {et~al.}(2018){McKernan}, {Ford}, {Bellovary}, {Leigh},
  {Haiman}, {Kocsis}, {Lyra}, {Mac Low}, {Metzger}, {O'Dowd}, {Endlich}, \&
  {Rosen}}]{2018ApJ...866...66M}
{McKernan}, B., {Ford}, K.~E.~S., {Bellovary}, J., {et~al.} 2018, \apj, 866, 66

\bibitem[{Miller {et~al.}(2020)Miller, Callister, \& Farr}]{Miller2020}
Miller, S., Callister, T.~A., \& Farr, W.~M. 2020, Astrophys. J., 895, 128

\bibitem[{Millman \& Aivazis(2011)}]{pythonForScientists}
Millman, K.~J., \& Aivazis, M. 2011, Computing in Science Engineering, 13, 9

\bibitem[{{Miralda-Escud{\'e}} \& {Gould}(2000)}]{2000ApJ...545..847M}
{Miralda-Escud{\'e}}, J., \& {Gould}, A. 2000, \apj, 545, 847

\bibitem[{{Morris}(1993)}]{1993ApJ...408..496M}
{Morris}, M. 1993, \apj, 408, 496

\bibitem[{{Olejak} {et~al.}(2020){Olejak}, {Fishbach}, {Belczynski}, {Holz},
  {Lasota}, {Miller}, \& {Bulik}}]{2020arXiv200411866O}
{Olejak}, A., {Fishbach}, M., {Belczynski}, K., {et~al.} 2020, \apjl, 901, L39

\bibitem[{Oliphant(2007)}]{pythonForScientificComputing}
Oliphant, T.~E. 2007, Computing in Science Engineering, 9, 10

\bibitem[{pandas~development team(2020)}]{pandas}
pandas~development team, T. 2020, pandas-dev/pandas: Pandas, vlatest,  Zenodo,
  doi:10.5281/zenodo.3509134.
\newblock \url{https://doi.org/10.5281/zenodo.3509134}

\bibitem[{Pratten {et~al.}(2020)Pratten, Husa, Garcia-Quiros, Colleoni,
  Ramos-Buades, Estelles, \& Jaume}]{Pratten:2020fqn}
Pratten, G., Husa, S., Garcia-Quiros, C., {et~al.} 2020, Phys. Rev. D, 102,
  064001

\bibitem[{Pratten {et~al.}(2021)}]{Pratten:2020ceb}
Pratten, G., {et~al.} 2021, Phys. Rev. D, 103, 104056

\bibitem[{Rodriguez {et~al.}(2018)Rodriguez, Amaro-Seoane, Chatterjee, \&
  Rasio}]{Rodriguez2018}
Rodriguez, C.~L., Amaro-Seoane, P., Chatterjee, S., \& Rasio, F.~A. 2018, Phys.
  Rev. D, 120, 151101

\bibitem[{Romero-Shaw {et~al.}(2019)Romero-Shaw, Lasky, \&
  Thrane}]{gwtc_eccentricity}
Romero-Shaw, I.~M., Lasky, P.~D., \& Thrane, E. 2019, MNRAS, 490, 5210

\bibitem[{{Romero-Shaw} {et~al.}(2021){Romero-Shaw}, {Lasky}, \&
  {Thrane}}]{gwtc2_eccentricity}
{Romero-Shaw}, I.~M., {Lasky}, P.~D., \& {Thrane}, E. 2021, arXiv e-prints,
  arXiv:2108.01284

\bibitem[{{Romero-Shaw} {et~al.}(2020){Romero-Shaw}, {Talbot}, {Biscoveanu},
  {D'Emilio}, \& et~al.}]{Romero-Shaw:2020:MNRAS}
{Romero-Shaw}, I.~M., {Talbot}, C., {Biscoveanu}, S., {D'Emilio}, V., \& et~al.
  2020, \mnras, 499, 3295

\bibitem[{Roulet {et~al.}(2021)Roulet, Chia, Olsen, Dai, Venumadhav, Zackay, \&
  Zaldarriaga}]{Roulet}
Roulet, J., Chia, H.~S., Olsen, S., {et~al.} 2021, Phys. Rev. D, 104, 083010

\bibitem[{Samsing(2018)}]{Samsing2018}
Samsing, J. 2018, Phys. Rev. D, 97, 103014

\bibitem[{{Samsing} {et~al.}(2020){Samsing}, {Bartos}, {D'Orazio}, {Haiman},
  {Kocsis}, {Leigh}, {Liu}, {Pessah}, \& {Tagawa}}]{2020arXiv201009765S}
{Samsing}, J., {Bartos}, I., {D'Orazio}, D.~J., {et~al.} 2020, arXiv e-prints,
  arXiv:2010.09765

\bibitem[{Schmidt {et~al.}(2012)Schmidt, Hannam, \& Husa}]{Schmidt2012}
Schmidt, P., Hannam, M., \& Husa, S. 2012, Phys. Rev. D, 86, 104063

\bibitem[{{Secunda} {et~al.}(2020){Secunda}, {Bellovary}, {Mac Low}, {Ford},
  {McKernan}, {Leigh}, {Lyra}, {S{\'a}ndor}, \& {Adorno}}]{2020ApJ...903..133S}
{Secunda}, A., {Bellovary}, J., {Mac Low}, M.-M., {et~al.} 2020, \apj, 903, 133

\bibitem[{{Sirko} \& {Goodman}(2003)}]{2003MNRAS.341..501S}
{Sirko}, E., \& {Goodman}, J. 2003, \mnras, 341, 501

\bibitem[{{Skilling}(2004)}]{skilling2004}
{Skilling}, J. 2004, in American Institute of Physics Conference Series, Vol.
  735, Bayesian Inference and Maximum Entropy Methods in Science and
  Engineering: 24th International Workshop on Bayesian Inference and Maximum
  Entropy Methods in Science and Engineering, ed. R.~{Fischer}, R.~{Preuss}, \&
  U.~V. {Toussaint}, 395--405

\bibitem[{{Skilling}(2006)}]{skilling2006}
{Skilling}, J. 2006, Bayesian Analysis, 1, 833.
\newblock \url{https://doi.org/10.1214/06-BA127}

\bibitem[{Smith \& Ashton(2021)}]{parallel_bilby}
Smith, R., \& Ashton, G. 2021, {Parallel Bilby: 1.0.1}, vv1.0.1,
  https://git.ligo.org.
\newblock \url{https://lscsoft.docs.ligo.org/parallel_bilby}

\bibitem[{{Smith} {et~al.}(2020){Smith}, {Ashton}, {Vajpeyi}, \&
  {Talbot}}]{pbilby_paper}
{Smith}, R. J.~E., {Ashton}, G., {Vajpeyi}, A., \& {Talbot}, C. 2020, \mnras,
  498, 4492

\bibitem[{{Speagle}(2020)}]{dynesty_paper}
{Speagle}, J.~S. 2020, \mnras, 493, 3132

\bibitem[{Stevenson {et~al.}(2017)Stevenson, Berry, \& Mandel}]{Stevenson}
Stevenson, S., Berry, C. P.~L., \& Mandel, I. 2017, MNRAS, 471, 2801

\bibitem[{{Stone} {et~al.}(2017){Stone}, {Metzger}, \&
  {Haiman}}]{2017MNRAS.464..946S}
{Stone}, N.~C., {Metzger}, B.~D., \& {Haiman}, Z. 2017, \mnras, 464, 946

\bibitem[{{Tagawa} {et~al.}(2020{\natexlab{a}}){Tagawa}, {Haiman}, {Bartos}, \&
  {Kocsis}}]{Tagawa:2020:ApJ}
{Tagawa}, H., {Haiman}, Z., {Bartos}, I., \& {Kocsis}, B. 2020{\natexlab{a}},
  \apj, 899, 26

\bibitem[{{Tagawa} {et~al.}(2020{\natexlab{b}}){Tagawa}, {Haiman}, \&
  {Kocsis}}]{2020ApJ...898...25T}
{Tagawa}, H., {Haiman}, Z., \& {Kocsis}, B. 2020{\natexlab{b}}, \apj, 898, 25

\bibitem[{{Talbot} {et~al.}(2019){Talbot}, {Smith}, {Thrane}, \&
  {Poole}}]{gwpopulation}
{Talbot}, C., {Smith}, R., {Thrane}, E., \& {Poole}, G.~B. 2019, \prd, 100,
  043030

\bibitem[{Talbot \& Thrane(2017)}]{spin_population_models}
Talbot, C., \& Thrane, E. 2017, Phys. Rev. D, 96, 023012

\bibitem[{{Talbot} \& {Thrane}(2018)}]{mass_population_models}
{Talbot}, C., \& {Thrane}, E. 2018, \apj, 856, 173

\bibitem[{{Tanaka} {et~al.}(2002){Tanaka}, {Takeuchi}, \&
  {Ward}}]{2002ApJ...565.1257T}
{Tanaka}, H., {Takeuchi}, T., \& {Ward}, W.~R. 2002, \apj, 565, 1257.
\newblock \url{https://iopscience.iop.org/article/10.1086/324713/pdf}

\bibitem[{{Thompson} {et~al.}(2005){Thompson}, {Quataert}, \&
  {Murray}}]{2005ApJ...630..167T}
{Thompson}, T.~A., {Quataert}, E., \& {Murray}, N. 2005, \apj, 630, 167

\bibitem[{Virtanen {et~al.}(2020)Virtanen, Gommers, Oliphant, Haberland, Reddy,
  Cournapeau, Burovski, Peterson, Weckesser, Bright, {van der Walt}, Brett,
  Wilson, Millman, Mayorov, Nelson, Jones, Kern, Larson, Carey, Polat, Feng,
  Moore, {VanderPlas}, Laxalde, Perktold, Cimrman, Henriksen, Quintero, Harris,
  Archibald, Ribeiro, Pedregosa, {van Mulbregt}, \& {SciPy 1.0
  Contributors}}]{SciPy}
Virtanen, P., Gommers, R., Oliphant, T.~E., {et~al.} 2020, Nature Methods, 17,
  261

\bibitem[{{Wang} {et~al.}(2021{\natexlab{a}}){Wang}, {McKernan}, {Ford},
  {Perna}, {Leigh}, \& {Mac Low}}]{2021arXiv211003698W}
{Wang}, Y., {McKernan}, B., {Ford}, S., {et~al.} 2021{\natexlab{a}}, arXiv
  e-prints, arXiv:2110.03698

\bibitem[{{Wang} {et~al.}(2021{\natexlab{b}}){Wang}, {Fan}, {Tang}, {Qin}, \&
  {Wei}}]{2021arXiv211010838W}
{Wang}, Y.-Z., {Fan}, Y.-Z., {Tang}, S.-P., {Qin}, Y., \& {Wei}, D.-M.
  2021{\natexlab{b}}, arXiv e-prints, arXiv:2110.10838

\bibitem[{Wysocki {et~al.}(2019)Wysocki, Lange, \& O'Shaughnessy}]{Wysocki2019}
Wysocki, D., Lange, J., \& O'Shaughnessy, R. 2019, Phys. Rev. D, 100, 043012

\bibitem[{{Yang} {et~al.}(2019{\natexlab{a}}){Yang}, {Bartos}, {Haiman},
  {Kocsis}, {M{\'a}rka}, {Stone}, \& {M{\'a}rka}}]{2019ApJ...876..122Y}
{Yang}, Y., {Bartos}, I., {Haiman}, Z., {et~al.} 2019{\natexlab{a}}, \apj, 876,
  122

\bibitem[{{Yang} {et~al.}(2020){Yang}, {Gayathri}, {Bartos}, {Haiman},
  {Safarzadeh}, \& {Tagawa}}]{2020ApJ...901L..34Y}
{Yang}, Y., {Gayathri}, V., {Bartos}, I., {et~al.} 2020, \apjl, 901, L34

\bibitem[{{Yang} {et~al.}(2019{\natexlab{b}}){Yang}, {Bartos}, {Gayathri},
  {Ford}, {Haiman}, {Klimenko}, {Kocsis}, {M{\'a}rka}, {M{\'a}rka}, {McKernan},
  \& {O'Shaughnessy}}]{2019PhRvL.123r1101Y}
{Yang}, Y., {Bartos}, I., {Gayathri}, V., {et~al.} 2019{\natexlab{b}}, \prl,
  123, 181101

\bibitem[{Zevin {et~al.}(2021)Zevin, Romero-Shaw, Kremer, Thrane, \&
  Lasky}]{eccentric_implications}
Zevin, M., Romero-Shaw, I.~M., Kremer, K., Thrane, E., \& Lasky, P.~D. 2021,
  Accepted for publication in Astrophys. J. Lett., arxiv/2106.09042

\end{thebibliography}

\end{document}